\documentclass[%
 reprint,
 amsmath,amssymb,
 aps,
]{revtex4-2}

\usepackage{graphicx}
\usepackage{dcolumn}
\usepackage{bm}

\begin{document}

\preprint{APS/123-QED}

\title{Super light-by-light scattering in vacuum induced by intense vortex lasers}

\author{Zhigang Bu$^{1}$}
\author{Lingang Zhang$^{1}$}
\author{Shiyu Liu$^{1}$}
\author{Baifei Shen$^{2}$}
\author{Ruxin Li$^{1,3}$}
\author{Igor P. Ivanov$^{4,*}$}
\author{Liangliang Ji$^{1,\dag}$}
\affiliation{
 $^{1}$State Key Laboratory of Ultra-intense Laser Science and Technology, Shanghai Institute of Optics and Fine Mechanics (SIOM), Chinese Academy of Sciences (CAS), Shanghai 201800, China\\
 $^{2}$Department of Physics, Shanghai Normal University, Shanghai 200234, China\\
 $^{3}$Shanghai Tech University, Shanghai 201210, China\\
 $^{4}$School of Physics and Astronomy, Sun Yat-sen University, Zhuhai 519082, China
}%

\date{\today}

\begin{abstract}
Collision of ultra-intense optical laser and X-ray free electron laser (XFEL) pulses is a promising approach to detecting nonlinear vacuum polarization (VP), a long-standing prediction of quantum electrodynamics remaining to be tested. Identifying the signals induced by polarized vacuum relies on purifying the X-ray polarization and poses significant challenges due to strongly reduced signal and low signal-to-noise ratio (SNR). Here we propose an approach that allows one to directly detect VP signals without the need for an X-ray polarizer. We discover a super light-by-light scattering (super-LBL) effect in collision of an X-ray probe with an intense laser in vortex modes, through which signal photons are kicked out of the X-ray background with large tangential momentum. Super-LBL originates from the gradient force of the vortical vacuum source in azimuthal direction and induces momentum exchange beyond the transverse momentum of laser-photon. This effect efficiently sets the scattered signal photons apart from the X-ray background, producing observable signals with both the strength and SNR about two orders of magnitude higher than those from the known VP effects. This finding paves the way for single-shot detection of nonlinear VP phenomena with current ultra-intense laser and XFEL technologies.
\end{abstract}

\maketitle


\textit{Introduction}---According to Quantum Electrodynamics (QED), vacuum is filled with quantum fluctuations of virtual positron-electron pairs. Through these fluctuations, photon-photon interactions can occur even in vacuum. For this reason, vacuum exhibits weak anisotropic properties in the presence of strong electromagnetic fields and influences the photon propagation. This is known as nonlinear QED vacuum polarization (VP), first predicted by Heisenberg and Euler in 1936 \cite{r01}. VP effects involving virtual photons have been observed, for example light-by-light scattering (LBL) in heavy-ion collisions \cite{rv01}, vacuum birefringent signals in neutron star observation \cite{rv02}, Delbr\"{u}ck scattering \cite{rv03} and photon splitting \cite{rv04} in nuclear Coulomb fields. Also, the detection of cosmic $\gamma$-ray horizon \cite{rv05}  provides potential cosmic evidence for the scattering of $\gamma$-photons with extragalactic background light or cosmic microwave background below the pair-production threshold. However, real-photon VP has never been detected in laboratory. Ultra-intense lasers open up possibilities for inducing and detecting real-photon processes, such as vacuum birefringence \cite{r02,r03,r04}, LBL \cite{r05,r06}, vacuum diffraction \cite{r07,r08}, and generation of high-order harmonics and electromagnetic shock wave in vacuum \cite{r09,r10}. Their signal strengths scale as $\sim(I_{L}/I_{cr})^{2}$, where $I_{cr}=2.3\times10^{29}\:\textrm{W/cm}^{2}$ is the intensity of Schwinger critical field \cite{r11} and $I_{L}$ is the intensity of the drive laser field.

With the present-day laser technology, the laser intensity achieved in laboratory has exceeded $10^{22}\:\textrm{W/cm}^{2}$ \cite{r12,r13,r14,r15} and is poised to reach $10^{23}\:\textrm{W/cm}^{2}$ \cite{r16}. At the same time, X-ray Free Electron Laser (XFEL) facilities deliver high-flux, high-quality X-ray beams. It is increasingly promising to detect real-photon QED VP by employing an ultra-intense laser as a driver and an X-ray as a probe \cite{r04,r16}. Nonetheless, with current laser technologies, the total VP signal strength is at least ten orders of magnitude lower than the X-ray probe background \cite{r04,r16,r17}. One is forced to produce enough signal photons and extract them from the X-ray background to significantly improve the signal-to-noise ratio (SNR).

There are two potential schemes to detect laser-driven VP effects. One is to rely on the vacuum birefringence effect \cite{r04,r18} and to detect polarization-flipped signal photons with the aid of an X-ray polarizer. As no more than 7\% of the scattered photons are polarization-flipped \cite{r17}, the birefringent signal photon count, even for $I_{L}=10^{23}\:\textrm{W/cm}^{2}$, drops to about $10^{-11}$ relative to the X-ray probe. Moreover, a high-resolution polarizer relies on multiple reflections of the X-ray beam in the polarizer crystal \cite{r19,r20}. This places strict requirements on the beam bandwidth and divergence angle \cite{r19,r21} and, at the same time, compromises the polarizer transmission efficiency, further attenuating the detectable signal by one order of magnitude. Dark-field detection \cite{r22} provides another method for enhancing the SNR and reducing the detection accuracy of the polarizer. However, it also reduces the probe photon number by blocking the central part of the X-ray beam before its focus. Adjusting the collision angle and transverse impact parameter of the drive laser and X-ray probe increases the divergence angle of the signal photon above that of the X-ray background \cite{r18}, which may also relax the requirements for the detection accuracy of the X-ray polarizer.

The second scheme is to detect a signal from the LBL scattering, regardless of the photon polarization. The scattered photon number in LBL is about 14 times higher than that of vacuum birefringence \cite{r17}. In the two-beam collision configuration (a drive laser vs. a probe), the energies and momenta of the scattered photons are almost unchanged with only very small amount of signal photons deflected outside the probe light cone \cite{rv06}, so that polarization measurement is necessary under current conditions. This obstacle can be avoided in an oblique collision of a probe pulse with two drive laser pulses \cite{r06,r23,r24}: the signal photons acquire a small transverse momentum (twice the transverse momentum of the drive lasers \cite{r24} in a symmetric setting), diverge from the probe background, and can be detected. As the collision angle increases, the transverse momenta grow, but the signal strength drops by two orders of magnitude \cite{r17}. With additional challenges associated with experimental control of the three-beam collision, detection of LBL seems no easier than of vacuum birefringence.

Vortex drive lasers have also been proposed to improve SNR in LBL \cite{rv07,rv08,rv09}, showing modulated profiles and orbital-angular-momentum (OAM) features in the signals. Here, the detectable signal strength is still much less than the total scattered light, showing no obvious advantage over Gaussian laser drivers. Therefore, one faces the dilemma that while improving the SNR the detectable signal strength is substantially suppressed. Signal accumulation over multiple shots seems necessary, which imposes stringent requirements on the repetition rate of ultra-intense lasers. As a result, experimental detection of VP effects remains a major unresolved challenge.

In this work, we theoretically demonstrate that these challenges can be addressed in a novel LBL scattering effect, in which the nonlinear QED vacuum driven by an intense laser in a mixed vortex mode induces a local vortex phase effect and generates a surprisingly large tangential momentum kick to the scattered photons. We refer to this effect as the super light-by-light scattering (super-LBL). Super-LBL steers the scattered photons out of the X-ray background and generates an observable signal with an SNR exceeding 100. This new effect is based on the counter-propagating two-beam collision and avoids the signal suppression due to the large collision angle. Moreover, it does not depend on the X-ray polarization, eliminating the need for polarizers and the associated signal depletion. Thus, super-LBL can produce more observable signal photons than the ordinary VP effects. We further demonstrate in a particle-in-cell (PIC) simulation that the drive laser in the mixed vortex mode can be generated using a custom-tailored double-ring spiral phase plate with the current experimental technologies. Our findings have significant advantages for detecting nonlinear QED VP effects, suggesting a promising experimental scheme with the present-day ultra-intense laser and XFEL technologies, and even allowing for single-shot measurements.

\begin{figure}[t]
\includegraphics[width=8.6cm]{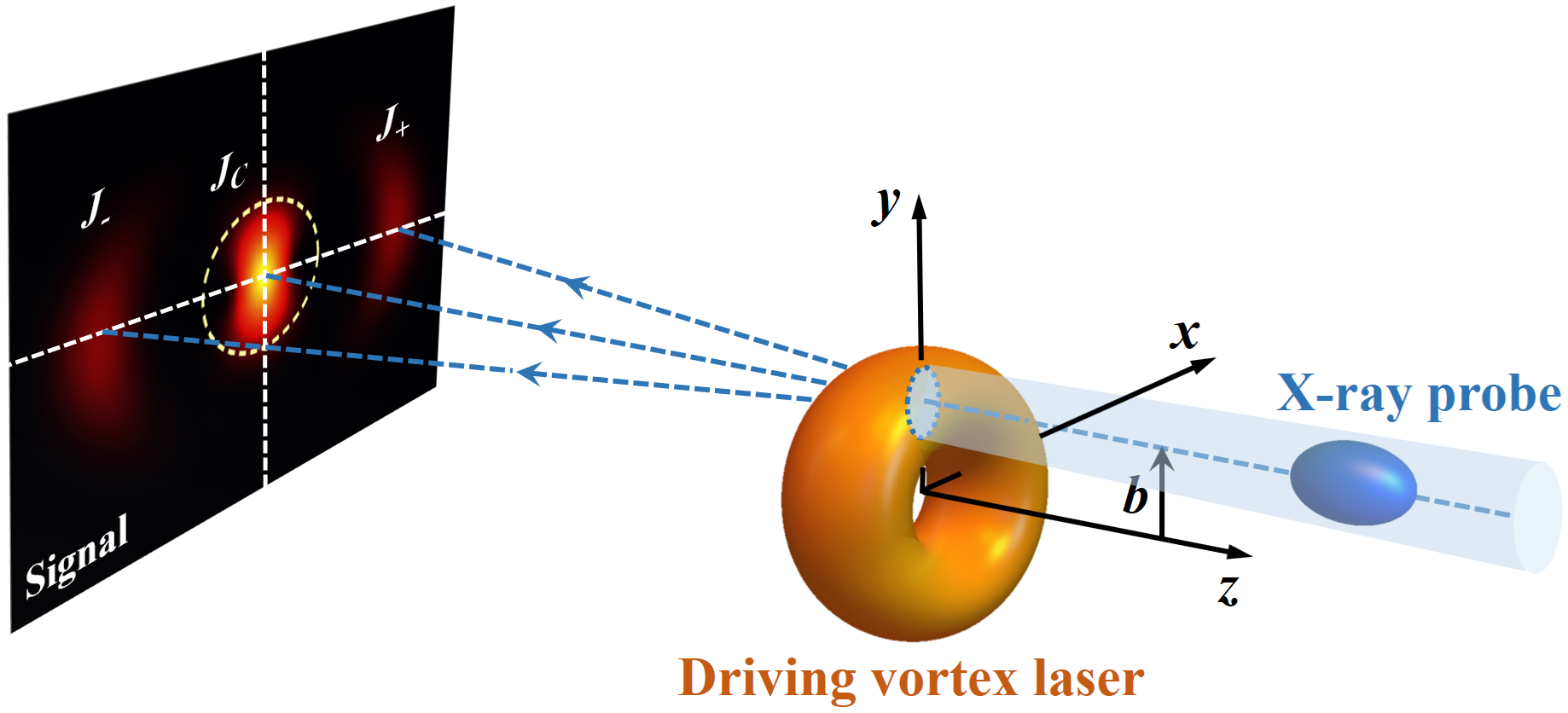}
\caption{Schematic layout of super-LBL scattering. An X-ray probe collides head-on, at the transverse impact parameter $\bm{b}$, with a drive laser pulse in a mixed vortex mode. The drive laser polarizes vacuum to generate the vortex vacuum source and induces photon-photon interaction. X-ray photons are scattered into signal photons by acquiring tangential momenta in azimuthal direction due to the local phase gradient force of the vortical vacuum source. Kicked out of the probe X-ray background (the yellow circle region on the screen), they form two detectable side-lobe signals.}
\label{fig:01}
\end{figure}

\textit{Theoretical model and demonstration}---Nonlinear QED VP can be analyzed using Heisenberg-Euler Lagrangian density \cite{r01,r11}, $\mathcal{L}=\mathcal{F}+2\xi\mathcal{F}^{2}+(7\xi/2)\mathcal{G}^{2}$, where the two Lorentz invariants of electromagnetic field are $\mathcal{F}=(\bm{E}^{2}-\bm{B}^{2})/2$ and $\mathcal{G}=\bm{E}\cdot\bm{B}$. Here, $\bm{E}$ and $\bm{B}$ are the total electric and magnetic fields, $\xi=\alpha/(45\pi E_{cr}^{2})$ is the VP coefficient, $\alpha=1/137$ is the fine structure constant, $E_{cr}=1.3\times 10^{16}\:\textrm{V/cm}$ is the Schwinger critical field strength. In the interaction of an intense laser pulse with an X-ray probe, the polarization and magnetization vectors, $\bm{P}$ and $\bm{M}$, are induced in vacuum and generate the equivalent charge and current densities, $\rho_{V}=\nabla\cdot\bm{P}$ and $\bm{J}_{V}=\partial_{t}\bm{P}+\nabla\times\bm{M}$. The vacuum charge and current in turn influence the X-ray probe and generates a $2\to2$ scattering process: an X-ray photon absorbs a laser photon and subsequently scatters into a signal photon and a laser photon. From the Lagrangian density, the wave equation of the scattered light is obtained,
\begin{equation}
(\partial_{t}^{2}-\nabla^{2})\bm{E}_{S}(t,\bm{x})=4\pi\bm{f}_{V}(t,\bm{x})
\label{eq:01},
\end{equation}
where $\bm{f}_{V}(t,\bm{x})=\nabla\rho_{V}-\partial_{t}\bm{J}_{V}$ is the equivalent vacuum source. Different polarization directions of this source determine the signals from vacuum birefringence and LBL. It is known that the signal of ordinary LBL is difficult to measure directly in two-beam collision because the signal photons experience no momentum change. What can be detected is a birefringent signal with the aid of an X-ray polarizer or the LBL signals in three-beam collision, but in either case the weak signal will be further attenuated by at least two orders of magnitude. To boost the measurement efficiency and SNR, we propose a super-LBL mechanism in the collision of an X-ray probe with a vortex drive laser. A vortex field possesses the helical phase factor $e^{il\theta}$, with $l$ being the OAM projection of each quantum of this field. The gradient of this vortex factor generates a local tangential momentum, $\bm{k}_{\theta}=-i\nabla_{\theta}e^{il\theta}=\frac{l}{r}\bm{e}_{\theta}$ \cite{r25}. A particle interacting with this vortex field is kicked in the azimuthal direction due to the transfer of the local momentum, which can exceed the transverse momentum of a quantum of the vortex field and is, therefore, referred to as the superkick \cite{r25,r26,r27}.

We find that the superkick effect also exists in the quantum vacuum and leads to super-LBL scattering. Figure~\ref{fig:01} presents a schematic layout of this process. We consider an ultra-intense drive laser pulse prepared in a mixed vortex mode (a superposition of two Laguerre-Gaussian (LG) modes with the OAM numbers $l_{1}$ and $l_{2}$), an X-ray probe pulse collides with it head-on, at the transverse impact parameter $\bm{b}$. The drive laser propagates along the +z-direction, linearly polarized with polarization vector $(1/\sqrt{2})(\bm{e}_{x}+\bm{e}_{y})$. To facilitate the distinction between the birefringent photons and LBL photons in the signals, we consider a Gaussian X-ray probe linearly polarized in y-direction. The polarization angle between them is $\pi/4$. Then the vacuum source is derived if we consider the slowly varying pulse envelopes and ignore the longitudinal component of the laser field,
\begin{equation}
\bm{f}_{V}(t,\bm{x})=\frac{\xi\omega_{X}^{2}}{32\pi}j(t,\bm{x})e^{-i(\omega_{X}t+k_{X}z)}(-3\bm{e}_{x}+11\bm{e}_{y})+C.C.
\label{eq:02}
\end{equation}
The frequency and momentum of the X-ray are $\omega_{X}$ and $\bm{k}_{X}=(0,0,-k_{X})$. With the vacuum source \eqref{eq:02}, the wave equation \eqref{eq:01} has an analytical solution in momentum space, with the signal light field $\bm{E}_{S}(t,\bm{k}_{S})$. Then, the momentum distribution of the scattered photons normalized by the total energy of X-ray probe is derived, $dW(\bm{k}_{S})/d^{3}\bm{k}_{S}=|\bm{E}_{S}(t,\bm{k}_{S})|^{2}/\int d^{3}\bm{x}|\bm{E}_{X}(t,\bm{x})|^{2}$, where $\bm{E}_{X}(t,\bm{x})$ is the electric field of the X-ray probe.

The $\bm{e}_{x}$-component of the vacuum source, Eq.~\eqref{eq:02}, generates vacuum birefringent signals. The amplitude ratio of $\bm{e}_{x}$ and $\bm{e}_{y}$ components is 3/11, thus the number of birefringent photons accounts for at most 9/130 ($\approx7\%$) of the total scattered photons. $j(t,\bm{x})$ in Eq.~\eqref{eq:02} is a scalar function determined by the pulse envelopes and the OAM numbers of the drive laser, which can be decomposed into three components: $j(t,\bm{x})=j_{C}(t,\bm{x})+j_{+}(t,\bm{x})+j_{-}(t,\bm{x})$. $j_{C}(t,\bm{x})$ is vortex-independent; $j_{\pm}(t,\bm{x})$ are vortex-dependent and have the opposite vortex phases carrying the OAM numbers $\pm(l_{1}-l_{2})$. Then the induced vacuum source possesses, in addition to the main non-vortex component $\bm{f}_{C}$, two new vortex components, $\bm{f}_{+}$ and $\bm{f}_{-}$. They appear exclusively due to a non-zero overlap of the drive laser field with itself even if one transmits $\pm(l_{1}-l_{2})$ units of OAM from the drive field to an X-ray photon. These vortex components can be visualized as induced by absorption of a laser photon with the OAM $l_{1}$ (or $l_{2}$) and re-emitting it into a laser photon with the OAM $l_{2}$ (or $l_{1}$), see Supplemental Material for more information. They induce a local tangential momentum that is transferred to the scattered photons, and generate the super-LBL signals $\bm{E}_{\pm}(t,\bm{k}_{S})$. As it can be much larger than the transverse momentum exchange in the standard VP effect, the signal photons are kicked out of the X-ray background, generating a directly measurable signal.

\begin{figure}[t]
\includegraphics[width=8.6cm]{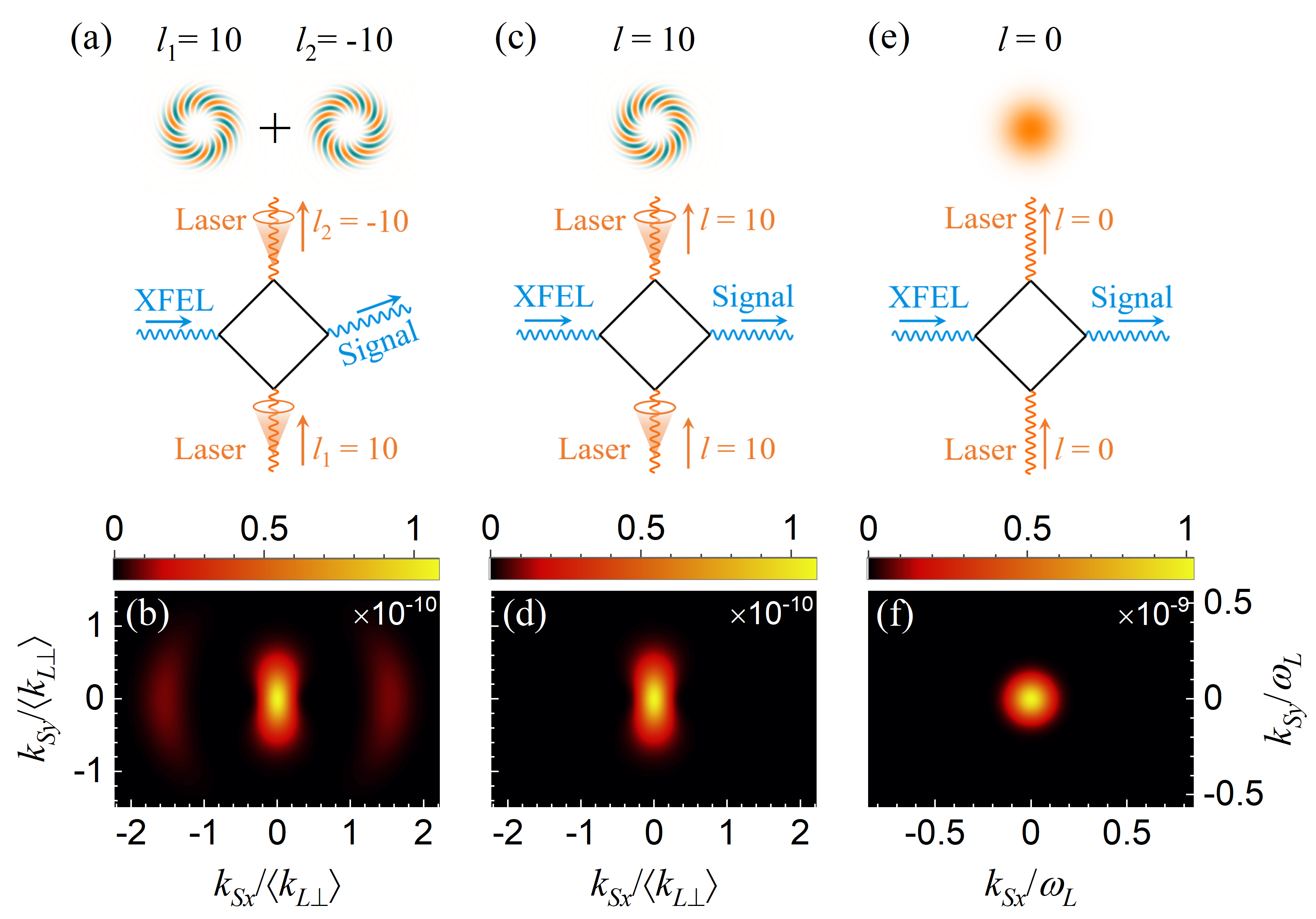}
\caption{Transverse momentum distributions of the signal photons. The drive laser is in coherent superposition mode of two LG components with OAM numbers $l_{1}=-l_{2}=10$ (a), single LG mode with OAM number $l=10$ (c) and Gaussian mode (e), the scattering Feynman diagrams are also shown in these figures. (b), (d) and (f) The corresponding transverse momentum distributions of the scattered photons, $dW(\bm{k}_{S\bot})$.}
\label{fig:02}
\end{figure}

The super-LBL mechanism is demonstrated in Fig.~\ref{fig:02}. We use the parameters accessible in the 100PW-class laser facility-the Station-of-Extreme-Light (SEL) at SHINE \cite{r16}. The wavelength of the drive laser is 910\,nm, total pulse energy is $1500\,\textrm{J}$, and duration of full width at half maximum (FWHM) is $15\,\textrm{fs}$, respectively. The drive laser is prepared in a superposition mode of two LG components with OAM numbers $l_{1}=-l_{2}=10$. Here the peak-to-peak diameter of the LG focal spot is $10\,\mu\textrm{m}$, corresponding to peak intensity of $5.9\times10^{23}\textrm{W}/\textrm{cm}^{2}$. The photon energy of the X-ray probe is $10\,\textrm{keV}$, FWHM duration $22.5\,\textrm{fs}$, and FWHM focal spot $3.5\,\mu\textrm{m}$. We set the impact parameter along the y-direction with $b_{y}=5\,\mu\textrm{m}$. Figure~\ref{fig:02}(a) shows the laser modes and Feynman diagram of the super-LBL. Figure~\ref{fig:02}(b) presents the transverse momentum distribution of the scattered photons, $dW(\bm{k}_{S\bot})/d^{2}\bm{k}_{S\bot}$. The central bright spot corresponds to the ordinary LBL scattering, while the two side lobes displaced in the direction perpendicular to $\bm{b}$ represent the hallmark signal of super-LBL. Compared to the X-ray probe background, the momenta of LBL signals hardly changed, their distribution showing only slight broadening in the y-direction. However, their strength is ten orders of magnitude lower than that of the X-ray probe. Therefore, the LBL signal cannot be extracted from X-ray background, unless an X-ray polarizer is used to detect the 7\% birefringent signal. On the contrary, the super-LBL signal acquires a tangential momentum in the x-direction, which is nearly twice as much as the average transverse momentum of the drive LG laser $\langle k_{L\bot}\rangle$. This allows the signal to be clearly offset from the X-ray background with a high SNR. For comparison, we show the scattering of an X-ray probe with a drive laser in a single-LG mode carrying OAM $l=10$ (Fig.~\ref{fig:02}(c) and (d)) and with a Gaussian laser pulse (Fig.~\ref{fig:02}(e) and (f)). The FWHM waist of the Gaussian laser pulse is $5\,\mu\textrm{m}$ following the designed parameter in SEL. The peak intensities of the single-LG pulse and Gaussian pulse are thus $3\times10^{23}\textrm{W}/\textrm{cm}^{2}$ and $6.5\times10^{23}\textrm{W}/\textrm{cm}^{2}$ with the same 1500\,J pulse energy. In both cases, the transverse momentum distributions of the scattered photons feature only one central bright spot corresponding to LBL scattering.

\begin{figure}[t]
\includegraphics[width=8.6cm]{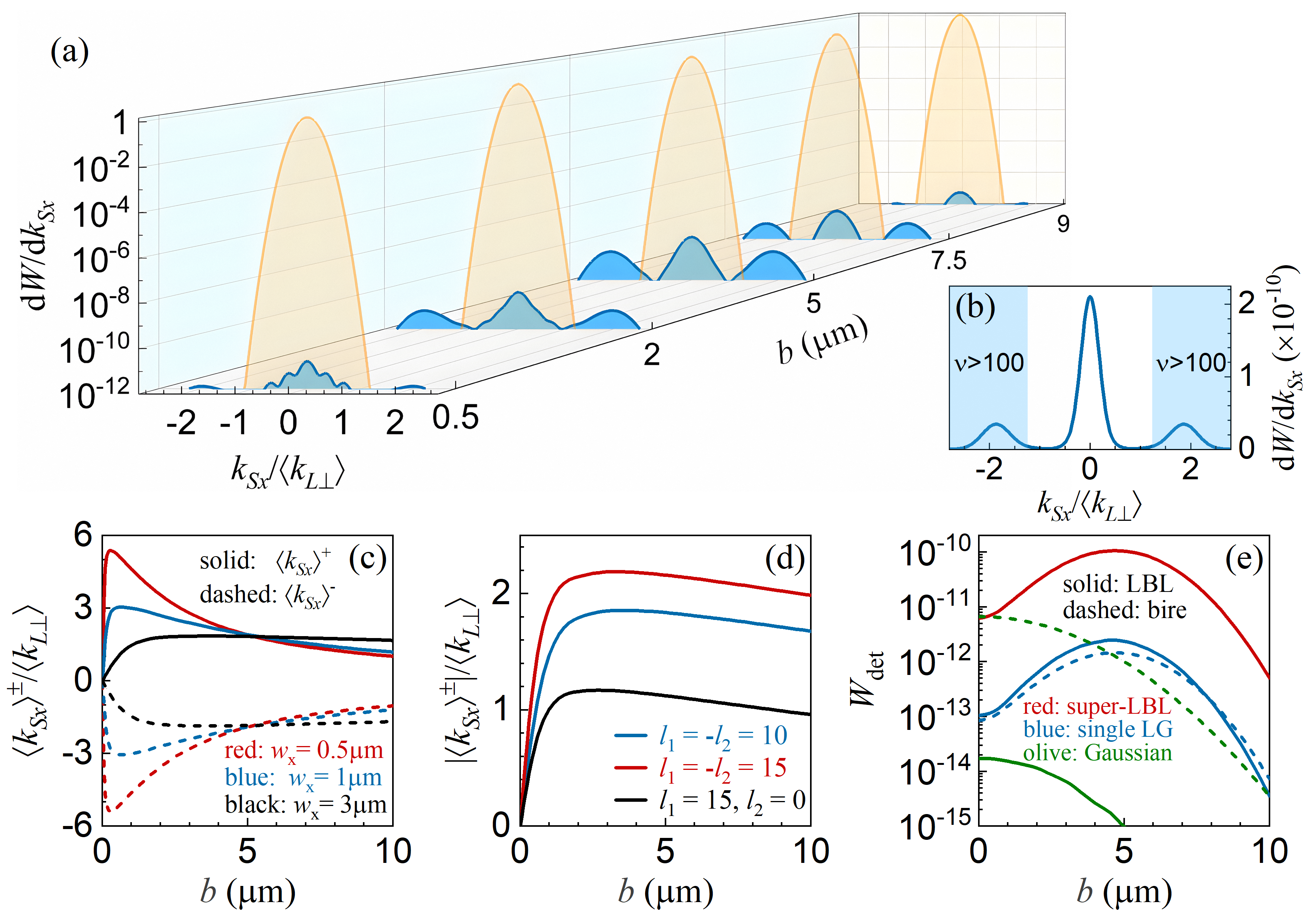}
\caption{Analysis of the observable signals. Tangential momentum distributions of signal photons for various impact parameters $\bm{b}$ (a) and for $b_{y}=5\mu\textrm{m}$ with SNR $\nu>100$ (blue shade area) (b). Average tangential momenta $\langle k_{Sx}\rangle^{\pm}$ of the super-LBL signals for various X-ray focal spot size (c) and for various vortex modes of drive laser. (e) Scattering probabilities of the detectable super-LBL signals, ordinary LBL signals and birefringent signals driven by single-LG laser and Gaussian laser versus the impact parameter $b$.}
\label{fig:03}
\end{figure}

\textit{Measurable signal analysis}---Turning to the question of feasibility of experimental observation of super-LBL, we need to address two issues. 1) Enough signal photons must be generated for the experimental detection. 2) The signal photons must receive sufficient tangential momentum to be offset from the X-ray background and produce a high SNR, $\nu=N_{X,total}dW(\bm{k}_{S})/dN_{X}(\bm{k}_{X})$, where $N_{X}(\bm{k}_{X})$ and $N_{X,total}$ are the momentum distribution and the total photon number of the X-ray probe. By performing the $dk_{Sy}dk_{Sz}$ integration of $dW(\bm{k}_{S})$, we compare the tangential momentum distribution of the scattered photons with the X-ray background for various impact parameters $b$ in Fig.~\ref{fig:03}(a). The central peaks of the blue lines are the ordinary LBL signals, and the two side peaks are the super-LBL signals. We find that the LBL signals are obscured by the X-ray background (orange areas), while the super-LBL signals are kicked out and distinctly visible. The signals are strongest when the impact parameter is around $b_{y}=5\:\mu\textrm{m}$, which corresponds to the focal spot size of the drive laser. Figure~\ref{fig:03}(b) shows the tangential momentum distribution of the scattered photons for $b_{y}=5\:\mu\textrm{m}$, where the blue shading is the area with SNR $\nu>100$. This indicates that the SNR of the super-LBL signals is very large.

We calculate the average value of the tangential momentum of the super-LBL signal photons as $\langle\bm{k}_{S\bot}\rangle^{\pm}=\int d^{3}\bm{k}_{S}\bm{k}_{S\bot}|\bm{E}_{\pm}(\bm{k}_{S})|^{2}/\int d^{3}\bm{k}_{S}|\bm{E}_{\pm}(\bm{k}_{S})|^{2}$. Since the impact parameter $\bm{b}$ is along the y-axis, it can be demonstrated that $\langle\bm{k}_{S\bot}\rangle^{\pm}$ are directed along the x-axis: $\langle k_{Sy}\rangle^{\pm}=0$. Figure~\ref{fig:03}(c) shows $\langle k_{Sx}\rangle^{\pm}$ of the super-LBL signals for various X-ray focal spots normalized to the average transverse momentum of the drive LG laser $\langle k_{L\bot}\rangle$. The ``$\pm$'' modes are symmetrically distributed, and $\langle k_{Sx}\rangle^{\pm}$ can be several times larger than $\langle k_{L\bot}\rangle$, the feature that justifies the term superkick. Roughly, $\langle k_{Sx}\rangle^{\pm}$ is inversely proportional to the impact parameter $\bm{b}$. But as $\bm{b}$ approaches 0, $\langle k_{Sx}\rangle^{\pm}$ drops rapidly \cite{r25}. Considering the size of the laser focal spot, the tangential momentum is an average result of the phase tangential gradient over the overlapping area of focal spots of drive laser and X-ray probe. This means the signal photons could sense the local phase gradient force more accurately and gain larger tangential momentum if the X-ray focal spot is smaller. We also compare in Fig.~\ref{fig:03}(d) the magnitude of $|\langle k_{Sx}\rangle^{\pm}|$ in different drive LG modes and confirm its proportionality to the OAM number of the vacuum source, $\Delta l=\pm(l_{1}-l_{2})$.

The scattering probability of detectable signal photons is calculated by integrating the momentum distribution with sufficient SNR, $W_{det}=\int dW_{\nu>100}(\bm{k}_{S})$. In Fig.~\ref{fig:03}(e), we show the detectable probability of super-LBL signals as a function of the impact parameter, see the red solid line. The peak value reaches $1.1\times10^{-10}$ at $b_{y}=5\:\mu\textrm{m}$. If the X-ray probe contains $N_{X}=10^{12}$ photons, more than 100 signal photons are kicked out and experimentally detectable with SNR $\nu>100$. The FWHM is about $3.8\mu\textrm{m}$, which determines the transverse alignment accuracy goal of the two lasers in experiment. For comparison, we calculate the detectable probabilities with SNR $\nu>$1 of ordinary LBL driven by single-LG laser (blue solid line) and Gaussian laser (olive solid line), under the same pulse energy, and their corresponding birefringent signals with 10\% transmission efficiency of the X-ray polarizer \cite{r20} (dashed lines), see Fig.~\ref{fig:03}(e). The laser parameters are the same as in Fig.~\ref{fig:02}. As can be seen, these signals are more than one order of magnitude weaker than super-LBL signals. So super-LBL can generate enough detectable signals with a high SNR in the collision of an ultra-intense laser with an X-ray pulse, for example the 100 PW laser and 10 keV XFEL in SEL \cite{r16}. This effect opens up new opportunities for detecting VP effects with the present-day intense laser and XFEL technologies. It even allows for single-shot detection.

\begin{figure}[t]
\includegraphics[width=8.6cm]{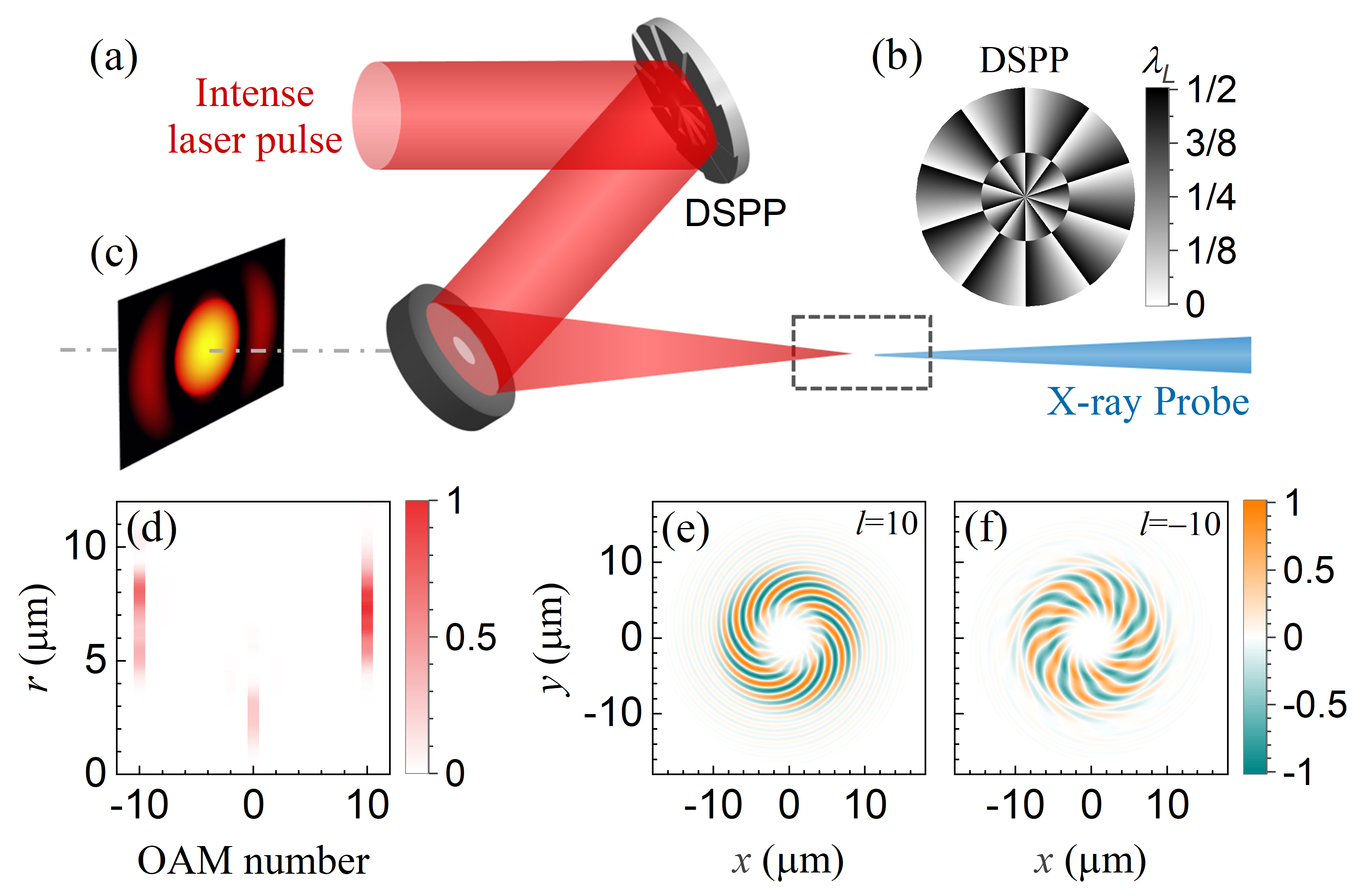}
\caption{(a) The proposed experimental scheme for detecting the super-LBL scattering. (b) The structure of DSPP. The step number in both the inner and outer circles of the DSPP is 10, but their directions are opposite. (c) The transverse distribution of the signal photons and the X-ray probe background after the vortex-laser-XFEL interaction. (d) The OAM spectrum of the reflected laser field. (e) and (f) The electric fields $E_{y}$ of the two vortex modes in transverse plane.}
\label{fig:04}
\end{figure}

\textit{Experimental scheme}---It remains to be shown how to produce an ultra-intense laser pulse in two-vortex mixed mode. We propose a feasible scheme presented in Fig.~\ref{fig:04}(a). An intense laser pulse interacts with a special designed double-ring spiral phase plate (DSPP) and is modulated into a two-vortex mixed mode. The structure of the DSPP is shown in Fig.~\ref{fig:04}(b), it is divided into the inner and outer circles, which may have different spiral steps to modulate the laser mode. The height of each step is half the laser wavelength, $H=\lambda_{L}/2$. In Fig.~\ref{fig:04}(b) the step number of the DSPP in both the inner and outer circles is 10, but their directions are opposite. When interacting with the DSPP, the laser pulse is converted into two vortex modes in inner and outer circles with OAMs $l_{1}=10$ and $l_{2}=-10$. After reflection from DSPP, the two vortex modes overlap during propagation, producing a mixed vortex laser pulse. The reflected laser is focused by a parabolic mirror and then collides with an X-ray probe to trigger super-LBL and generate the measurable signals, see Fig.~\ref{fig:04}(c). We confirmed the effectiveness of this DSPP-based modulating scheme through three-dimensional (3D) PIC simulation. A y-polarized incident laser propagates along z-axis, its wavelength is $\lambda_{L}=910\:\textrm{nm}$ and transverse envelope is super-Gaussian. Figure~\ref{fig:04}(d) shows the OAM spectrum of the reflected laser, which is dominated by $l=\pm10$ modes. The electric fields $E_{y}$ of these two modes in x-y plane are shown in Fig.~\ref{fig:04}(e) and (f). They possess nearly identical strengths and focal radii, indicating that the reflected laser is a coherent superposition of two OAM modes with $l_{1}=10$ and $l_{2}=-10$. Therefore, the simulation demonstrates the possibility of experimentally detecting super-LBL.

\textit{Conclusion and outlook}---We theoretically discovered a local vortex phase effect in LBL process, which generates super-LBL effect. When a compact X-ray probe collides with a drive laser pulse in a superposition mode of two vortex components, the induced vortical vacuum source generates a large local tangential momentum, which is transferred to the scattered photons and kicks them out of the X-ray probe background. The detecion of super-LBL does not rely on X-ray polarizers and the associated signal attenuation, avoids the complexity of the three-beam collision setting, and reduces the requirements for spatiotemporal alignment accuracy. With a 3D PIC simulation, we argue that ultra-intense two-vortex mixed laser necessary for super-LBL can be generated by using a DSPP. We find that super-LBL can generate enough signal photons with high SNR $>100$, both figures exceeding the vacuum birefringence and ordinary LBL counterparts by more than one order of magnitude. Not only do we propose this phenomenon as a means to detect the long-predicted nonlinear VP effects, but we argue that it can be done in a single shot experiment with the present-day intense laser and XFEL technologies. Super-LBL manifests as a super-oscillation phenomenon near the phase singularity \cite{r28}. Our study demonstrates that such super-oscillation can be induced in quantum vacuum. Realization of super-LBL will help verify QED processes in the extremely nonlinear regime and will reveal the quantum nature of the QED vacuum in an ultra-intense field.

~\\
\textit{Acknowledgments}---This work was supported by the National Science Foundation of China (Grant Nos. 12388102, 24G0011101), the Strategic Priority Research Program of Chinese Academy of Sciences (No. XDB0890303), the CAS Project for Young Scientists in Basic Research (Grant No. YSBR060), and the National Key R\&D Program of China (Grant No. 2022YFE0204800).

\nocite{*}
~\\
Corresponding author:\\
$^{*}$ivanov@mail.sysu.edu.cn\\
$^{\dag}$jill@siom.ac.cn

\bibliography{super-LBL}

\end{document}